\documentclass[twocolumn,aps,prl,showpacs,superscriptaddress,]{revtex4}
\usepackage{graphicx}

\begin{document}

\title{Nonequilibrium Spin Dynamics in a Trapped Fermi Gas with Effective
Spin-Orbit Interaction}
\author{Tudor D. Stanescu, Chuanwei Zhang, and Victor Galitski}
\affiliation{Condensed Matter Theory Center, Department of Physics, University of
Maryland, College Park, Maryland 20742-4111, USA}

\begin{abstract}
We consider a trapped atomic system in the presence of spatially
varying laser fields. The laser-atom interaction generates a
pseudospin degree of freedom (referred to simply as \textit{spin})
and leads to an effective spin-orbit  coupling for the fermions in
the trap. Reflections of the fermions from the trap boundaries
provide a physical mechanism for effective momentum relaxation and
non-trivial spin dynamics due to the emergent spin-orbit coupling.
We explicitly consider evolution of an initially spin-polarized
Fermi gas in a two-dimensional harmonic trap and derive
non-equilibrium behavior of the spin polarization. It shows
periodic echoes with a frequency equal to the harmonic trapping
frequency. Perturbations, such as an asymmetry of the trap, lead
to the suppression of the spin echo amplitudes. We discuss a
possible experimental setup to observe spin dynamics and provide
numerical estimates of relevant parameters.
\end{abstract}

\maketitle

Ultracold atomic gases have proven to be an ideal test-ground for the
experimental study of a variety of condensed matter phenomena \cite%
{Lewenstein}. A particularly interesting possibility is to realize
spin-orbit (SO) interaction in cold atomic systems. In atomic
gases, (pseudo)-spin represents a combination of different
hyperfine levels of atoms. It was proposed recently that motion of
atoms in position-dependent laser configurations may give rise to
an effective non-Abelian gauge
potential~\cite{oberg1,oberg2,Jaksch,Rusec,Osterloh,Zhu,Clark}. As
argued in this Letter, a similar setup may lead  to an effective
SO interaction. In particular, with proper laser configurations,
one can engineer Rashba or linear Dresselhaus SO coupling terms as
well as other types of couplings not accessible in solid state
systems. The resulting spin dynamics in the adiabatic regime are
remarkably simple, providing a robust experimental signature of
the underlying effective SO coupling. Throughout this article, we
refer to the pseudo-spin degree of freedom realized in cold-atom
systems as simply \textit{spin}.

One of the main aspects of spin dynamics in bulk condensed matter
systems is the Dyakonov-Perel spin relaxation~\cite{DyakPer}. This
mechanism involves random elastic scattering of electrons off of
impurities. These scatterings lead to spin relaxation, which is a
result of spin precession around a randomly oriented
momentum-dependent axis. Another example is an electron in a
quantum dot in the presence of SO coupling. In this case, the
existence of discrete energy levels results in oscillatory time
evolution of the spin polarization. The polarization decay and
relaxation to an equilibrium state occur only in the presence of
inelastic processes, such as electron-phonon
interactions~\cite{Khaets,Woods} or nuclear hyperfine
effects~\cite{Merkul,Semen}. In the absence of such processes, the
relaxation due to scatterings off of the quantum dot boundary is
qualitatively different from the bulk spin relaxation, as shown by
the semiclassical analysis, e.g. of Chang et al.~\cite{Chang}.

Atomic systems in the presence of spatially varying laser fields
offer the new remarkable possibility of observing strongly
non-equilibrium quantum spin dynamics in a many-particle system
without complications due to disorder that are always present in
condensed matter systems. In addition, by changing the geometry of
the trapping potential one can access various dynamic regimes
characterized by either regular or chaotic behavior of the spin
polarization~\cite{Milner,Friedman}. In this Letter we concentrate
on quantum spin dynamics of a Fermi gas confined in a
two-dimensional harmonic trap and in the presence of a simple SO
interaction term. We find that the spin polarization may show two
qualitatively different  behaviors dependent on the strength of
the SO coupling and on the number of particles. In the strong
coupling regime, the initially fully polarized Fermi gas becomes
completely unpolarized within a short time and remains unpolarized
most of the time. In the weak coupling regime, the spin
polarization never vanishes but oscillates with the period $T=2\pi
/\omega $, determined by the harmonic trapping frequency, $\omega
$, with the oscillation amplitudes dependent on the SO coupling
strength and the number of atoms. In both cases, spin polarization
exhibits periodic echoes with the echo frequency equal to the
harmonic trapping frequency. The echo amplitude and the period of
spin polarization oscillations are strongly modified in an
asymmetric elliptic harmonic trap.

\begin{figure}[tbp]
\begin{center}
\includegraphics[width=0.47\textwidth]{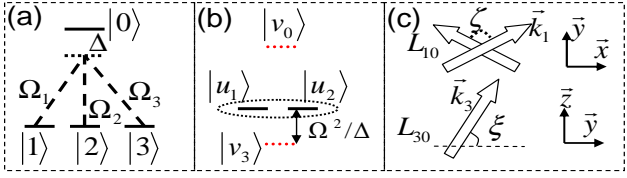}
\end{center}
\caption{Schematic representation of the experimental setup to
create effective SO interaction. (a) Coupling between the lasers
and the hyperfine states. (b) Energies of the degenerate dark
states ($|u_i\rangle$) and of the bright states ($|v_i\rangle$).
(c) Laser field configuration. } \label{Fig0}
\end{figure}

Consider an ultracold Fermi gas confined in a quasi-two dimensional ($xy$%
-plane) harmonic trap. Along the $z$ direction, the atomic dynamics is
\textquotedblleft frozen\textquotedblright\ by a high frequency optical trap
\cite{Raizen} or a deep optical lattice \cite{Spielman}.
The SO coupling may be generated using the tripod scheme shown in
Fig \ref{Fig0}(a), which is similar to the setup described in Ref.~\cite%
{Rusec}. The states $\left\vert 1\right\rangle $, $\left\vert
2\right\rangle $, and $\left\vert 3\right\rangle $ represent three
degenerate hyperfine ground states (e.g., different Zeeman
components of the hyperfine states of $^{6}$Li atoms). These
states are coupled to an excited state $|0\rangle $ by spatially
varying laser fields $L_{i0}$, with the corresponding Rabi
frequencies $\Omega _{1}$, $\Omega _{2}$, and $\Omega _{3} $. The
single-particle Hamiltonian is
\begin{equation}
\hat{\mathcal{H}}=\mathbf{p}^{2}/2m+\hat{V}_{\mathrm{trap}}+\hat{\mathcal{H}}%
_{a-l},  \label{Ham}
\end{equation}%
where $\mathbf{p}$ is the momentum, $\hat{V}_{\mathrm{trap}}=\sum_{j}V_{j}({%
\mathbf{r}})\left\vert j\right\rangle \left\langle j\right\vert $
represents the position-dependent trapping potential and
$\hat{\mathcal{H}}_{a-l}$ is the laser-atom interaction
Hamiltonian, given by $\hat{\mathcal{H}}_{a-l}=\Delta \left\vert
0\right\rangle \left\langle 0\right\vert - \left[ \Omega
_{1}\left\vert 0\right\rangle \left\langle 1\right\vert +\Omega
_{2}\left\vert 0\right\rangle \left\langle 2\right\vert +\Omega
_{3}\left\vert 0\right\rangle \left\langle 3\right\vert
+\mbox{h.c.}\right] $, where $\Delta $ is the detuning to the excited state $%
\left\vert 0\right\rangle $. The Rabi  frequencies can be
parameterized as $\Omega _{1}=\Omega \sin \theta \cos \phi
e^{iS_{1}}$, $\Omega _{2}=\Omega \sin \theta \sin \phi e^{iS_{2}}$, and $%
\Omega _{3}=\Omega \cos \theta e^{iS_{3}}$, with $\Omega
=\sqrt{|\Omega _{1}|^{2}+|\Omega _{2}|^{2}+|\Omega _{3}|^{2}}$. We
set $\hbar =1$.

The diagonalization of $\hat{\mathcal{H}}_{a-l}$ yields four eigenstates:
two degenerate dark states $\left\vert u_{1}\right\rangle =\sin \phi
~e^{-iS_{13}}\left\vert 1\right\rangle -\cos \phi ~e^{-iS_{23}}\left\vert
2\right\rangle $, $\left\vert u_{2}\right\rangle =\cos \theta \cos \phi
~e^{-iS_{13}}\left\vert 1\right\rangle +\cos \theta \sin \phi
~e^{-iS_{23}}\left\vert 2\right\rangle -\sin \phi \left\vert 3\right\rangle $%
, and two non-degenerate bright states $\left\vert
v_{0}\right\rangle \approx \left\vert 0\right\rangle $ and
$\left\vert v_{3}\right\rangle \approx \sin \theta \cos \phi
~e^{-iS_{13}}\left\vert 1\right\rangle +\sin \theta \sin \phi
~e^{-iS_{23}}\left\vert 2\right\rangle -\cos \phi \left\vert
3\right\rangle $, with $S_{ij}=S_{i}-S_{j}$ [Fig. \ref{Fig0}(b)].
Note that the energy of the dark states is not modified by the
laser field.
We assume $\omega \ll \Omega \ll \Delta $ and $\omega \ll \Omega
^{2}/\Delta $, where $\omega$ is the characteristic frequency of
the trapping potential, and neglect contributions of order $\Omega
/\Delta $ and smaller.
This condition implies that the states from the subspace spanned by $%
|u_{1}\rangle $ and $|u_{2}\rangle $ are well separated in energy
from the states $\left\vert v_{0}\right\rangle $ and $\left\vert
v_{3}\right\rangle $ , so that the coupling between dark and
bright states is negligible (adiabatic approximation). In the
present work we
concentrate on a particular configuration of laser fields [see Fig \ref{Fig0}%
(c)] characterized by $S_{1}=S_{2}$, $S_{31}\equiv S=mv_{s}y$ , $\phi
=mv_{\phi }x$, and the constant angle $\theta \in \lbrack 0,\pi ]$. The
laser field $L_{10}$ is generated by two laser beams propagating in the $xy$
plane and intersecting at an angle, $\zeta $. These two lasers form a
standing wave along the \thinspace $x$-direction and a plane wave along the $%
y$-direction. The laser field $L_{20}$ has the same configuration
except that the standing wave along the $x$-direction is shifted
in phase by $\pi /2 $. With such a setup, we have $\phi
=2k_{1}x\sin \left( \zeta /2\right) $ [or $mv_{\phi }=2k_{1}\sin
\left( \zeta /2\right) $] and $S_{1}=S_{2}=k_{1}y \cos \left(
\zeta /2\right) $, where $k_{1}=k_{2}$ are the wave-vectors for
the laser beams. Finally, the laser field $L_{30}$ is a plane wave
propagating in the $yz$ plane with an adjustable angle $\xi $ with
respect to the $y$-axis, leading to $S_{3}=k_{3}y\cos \xi $ [$
mv_{s}=k_{3}\cos \xi -k_{1}\cos \left( \zeta /2\right) $)].

The effective low-energy Hamiltonian is obtained by projecting the
Hamiltonian (\ref{Ham}) onto the subspace of the degenerate dark states $%
\left\vert u_{1}\right\rangle \otimes \left\vert u_{2}\right\rangle $~\cite%
{Rusec}
\begin{equation}
\hat{\mathcal{H}}_{u}=\left[ \frac{p^{2}}{2m}+w(r)\right] \hat{I}_{2}+\delta
_{0}\hat{\sigma}_{z}+\hat{\mathcal{H}}_{SO},  \label{Hueff}
\end{equation}%
where $\hat{I}_{2}$ represents the $2\times 2$ unit matrix and $\hat{\sigma}%
_{j} $ with $j\in \{x,y,z\}$ are the Pauli matrices.
Here $V_{1}=V_{2}=w(r)$ and $V_{3}=w(r)+\delta $ , where $w(r) =
m\omega ^{2}r^{2}/2$ (unless otherwise noted) is a symmetric
harmonic potential and $\delta $ is a constant shift for the
hyperfine state
$\left\vert 3\right\rangle $ that yields an effective Zeeman splitting $%
\delta _{0}=\sin ^{2}\theta \lbrack \delta -(v_{s}^{2}+v_{\phi
}^{2})/2]/2$ of the dark states and can be varied through
additional state-dependent laser fields.\cite{Grimm} The last term
in (\ref{Hueff}) is the SO coupling
\begin{equation}
\hat{\mathcal{H}}_{SO}=-v_{0}p_{x}\hat{\sigma}_{y}-v_{1}p_{y}\hat{\sigma}%
_{z},
\end{equation}
with $v_{0}=v_{\phi }\cos \theta $ and $v_{1}=v_{s} \sin ^{2}\theta /2$,
where the pseudo-spin represents the internal degree of freedom associated
with the degenerate dark states $\left\vert u_{1}\right\rangle $ and $%
\left\vert u_{2}\right\rangle $. We will call an atom in a $\left\vert
u_{1}\right\rangle $ ($\left\vert u_{2}\right\rangle $) state as a spin up
(down) particle. The term $\hat{\mathcal{H}}_{SO}$ of the effective
Hamiltonian couples this (pseudo)-spin degree of freedom to the motion of
the atom inside the trap.
The ``direction'' of the spin does not have any real space
significance and is solely determined by the choice of the basis
in the dark state subspace.

Atomic SO coupled systems open new possibilities to study strongly
non-equilibrium spin dynamics, in contrast to boundary linear
response effects usually considered in solid state systems with SO
coupling~\cite{CMSO}. Below, we study such a problem for the
simplest Ising-like spin orbit coupling, which already leads to a
non-trivial dynamics.  This choice corresponds to a constant phase
$S=const$; i.e., the plane-wave components of the three laser
fields $L_{i0}$ along $y$-direction share the same wave-vector. We
define the spin polarization $\mathcal{P}(t)$ as the difference
between the occupation numbers of the dark states $|u_1\rangle$
and $|u_2\rangle$. We assume that at $t=0$, the system is fully polarized $%
\mathcal{P}(0)=1$, i.e. all the particles are in the dark state $|u_1\rangle$%
. The corresponding single-particle quantum
mechanics problem can be easily solved exactly. If $\phi_{\alpha}(%
\mathbf{r})$ is an eigenstate of the operator $H_0=p^2/2m +w(\mathbf{r})$
with an eigenvalue $\epsilon_{\alpha}$, then the eigenfunction of the
SO coupled Hamiltonian $\hat{\mathcal{H}}_u=H_0\hat{I}_2- v_0 p_x%
\hat{\sigma}_y$ is
\begin{equation}
\overline{\psi}_{\alpha \lambda}(\mathbf{r}) = \phi_{\alpha}(\mathbf{r}%
)e^{i\lambda m v_0 x}\frac{1}{\sqrt{2}}\left(%
\begin{array}{c}
1 \\
i\lambda%
\end{array}
\right),  \label{psialf}
\end{equation}
with $\lambda=\pm1$. For a harmonic trap, $w(r)=m\omega^2r^2/2$, $%
\phi_{\alpha}(\mathbf{r})=\varphi_{n_x}(x)\varphi_{n_y}(y)$ can be written
as a product of the harmonic oscillator eigenfunctions $\varphi_n$ and the
energy spectrum becomes $\epsilon_{n_xn_y} = \omega(n_x+n_y+1)$. The
eigenfunctions (\ref{psialf}) are degenerate with respect to $\lambda$ and
have the eigenvalues $E_{\alpha}=\epsilon_{\alpha} - m v_0^2/2$.

\begin{figure}[tbp]
\begin{center}
\includegraphics[width=0.42\textwidth]{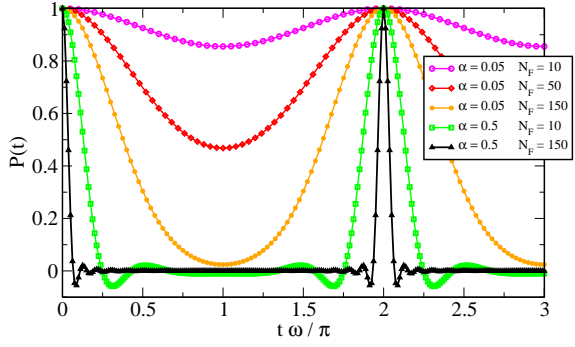}
\end{center}
\caption{(Color online) Spin polarization as a function of time.
Relaxation curves are plotted for $N=(N_F+1)(N_F+2)/2$ particles
in a harmonic trap and
in the presence of a (pseudo) SO interaction parameterized by $%
\protect\alpha=(mv_0^2/2\protect\omega)^{1/2}$. The $2\protect\pi/\protect%
\omega$ periodicity is due to the equal spacing between the
harmonic oscillator levels. When $\protect\alpha N_F^{1/2}$ is
large (black curve with triangles), the polarization is
characterized by fast relaxation followed by periodic echoes. }
\label{FCS1}
\end{figure}
Next, we define the spin polarization $\mathcal{P}(t) =
\frac{1}{N}\langle\Phi(t)|\hat{P}_z|\Phi(t)\rangle$ as an the
average over the quantum many-body state $\Phi(t)$ obtained
from the initial state $\Phi_0$ by applying the time evolution operator $%
\hat{\mathcal{U}}(t) = \exp(-i\hat{\mathcal{H}}t)$. The operator
$\hat{P}_z$ can be expressed in terms of field operators $\hat{\Psi}(%
\mathbf{r})$ as $\hat{P}_z = \int d^2r~\hat{\Psi}^{\dagger}({\mathbf{r}})\hat{\sigma}_z\hat{%
\Psi}({\mathbf{r}})$
In the Heisenberg representation, the polarization becomes $%
\hat{P}_z(t)
=\hat{\mathcal{U}}^{-1}(t)\hat{P}_z\hat{\mathcal{U}}(t)$. We can
obtain the explicit time dependence by expressing the field
operators in terms of creation and annihilation operators
associated with the single particle eigenstates (\ref{psialf}),
$\hat{\Psi}({\mathbf{r}}) =\sum_{\alpha,
\lambda}\overline{\psi}_{\alpha \lambda}({\mathbf{r}})
\hat{a}_{\alpha \lambda}$,
and taking advantage of the simple time dependence of these operators, $\hat{%
a}_{\alpha \lambda}(t) = \hat{a}_{\alpha
\lambda}~e^{-iE_{\alpha}t}$. In order to evaluate the action of
the polarization operator on $\Phi_0$, it is convenient to
introduce a set of operators $\hat{b}_{\alpha \sigma}$ associated
with the fully
polarized single particle states $\overline{\psi}_{\alpha \sigma}^{(0)}(%
\mathbf{r})=\phi_{\alpha}(\mathbf{r})\overline{\chi}_{\sigma}$, with $%
\overline{\chi}^{\dagger}_{\uparrow}=(1, 0)$ and $\overline{\chi}%
^{\dagger}_{\downarrow}=(0, 1)$ and satisfying the relation $\hat{b}_{\alpha
\sigma}=\sum_{\beta\lambda}\langle\overline{\psi}_{\alpha \sigma}^{(0)}|%
\overline{\psi}_{\beta\lambda} \rangle\hat{a}_{\alpha \lambda}$.
In terms of b-type operators, the initial state can be written as
$|\Phi_0\rangle = \prod_{\alpha}^\prime
\hat{b}^{\dagger}_{\alpha\uparrow}|\emptyset\rangle$ where
$|\emptyset\rangle$ represents the vacuum and the prime signifies
that the product is constrained by the condition $E_{\alpha}\leq
E_F$, where $E_F=\omega(N_F+1)$ is the ``Fermi'' energy. After
recasting $\hat{P}_z(t)$ in terms of b-type operators, the
polarization becomes
\begin{eqnarray}
\mathcal{P}(t) &=& \frac{1}{2N}\sum_{\sigma, \alpha, \beta }
\sum_{\gamma}^{(E_{\gamma}\leq E_F)}~\langle\phi_{\gamma}|e^{i\sigma
\xi}|\phi_{\alpha}\rangle  \label{Pfinal} \\
&\times&\langle\phi_{\alpha}|e^{-2i\sigma \xi}|\phi_{\beta}\rangle
\langle\phi_{\beta}|e^{i\sigma \xi}|\phi_{\gamma}\rangle
e^{i(\epsilon_{\alpha}-\epsilon_{\beta})t},  \nonumber
\end{eqnarray}
where $\xi=m v_0 x$ and $\sigma=\pm1$. Eq.~(\ref{Pfinal})
expresses the polarization of a non-interacting many-body system
with SO
interaction in terms of matrix elements of the single particle states $%
\phi_{\alpha}$ describing the motion of an atom in the confining
potential. The effects of the SO coupling are contained in the
parameter $\xi$
and the finite number of particles enters through the constraint on the $%
\gamma$ sum. Equation (\ref{Pfinal}) contains matrix elements
$\langle\varphi_{n}|e^{-2i\sigma \xi}|\varphi_{m}\rangle$ of the
harmonic eigenfunctions $\varphi_{n}(x)$ with energies
$\epsilon_n=\omega(n+1/2)$. The summations over the quantum
numbers are performed numerically. The time dependence of the
pseudo-spin polarization for $N$ particles that are initially in
the dark state $|u_1\rangle$ and occupy the first $N_F$ energy
levels is shown in Fig.~\ref{FCS1}. We parameterize the strength
of the SO coupling by $\alpha=(mv_0^2/2\hbar\omega)^{1/2}$. Notice
that all the curves in Fig.~\ref{FCS1} display a periodic
structure.

Spin relaxation depends strongly on both the strength of the SO
interaction (i.e. on $\alpha$) and the number of particles (i.e.
on $N_F$). In the weak coupling limit, $\alpha^2 N_F\ll 1$, the
polarization deviates slightly from unity and we have
\begin{equation}
\mathcal{P}(t) \approx 1-4\alpha^2\left(\frac{2N_F}{3}+1\right)[1-\cos(%
\omega t)].
\end{equation}
At strong coupling, $\alpha^2 N_F\gg 1$, the polarization is
characterized by fast spin relaxation to zero and periodic spin
echoes that restore $\mathcal{P}$ to its initial value. Performing
a power expansion of the polarization at small times we get
\begin{equation}
\mathcal{P}(t) \approx 1 - \sum_n c_n (\alpha\sqrt{N_F}\omega t)^{2n},
\label{Pexpand}
\end{equation}
where $c_n$ are numerical coefficients. From this we obtain the time scale
of the fast relaxation, $\Delta t = 1/(\omega\alpha\sqrt{N_F})$, which is
also the characteristic width of the echo peaks.

\begin{figure}[tbp]
\begin{center}
\includegraphics[width=0.42\textwidth]{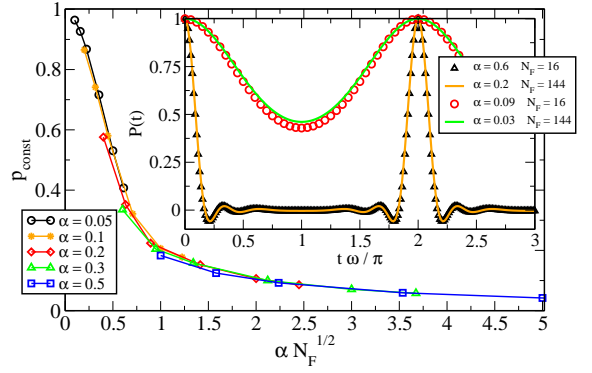}
\end{center}
\caption{(Color online) The time independent contribution to the
polarization as a function of the effective coupling parameter $\protect%
\alpha N_F^{1/2}$. For each $\protect\alpha$, $P_{\mathrm{const}}$
was
determined for several values of $N_F$. In the strong coupling limit $P_{%
\mathrm{const}}$ scales with $\protect\alpha N_F^{1/2}$. Inset:
relaxation curves for sets of parameters corresponding to the same
effective coupling: $\protect\alpha N_F^{1/2}=2.4$ (black
triangles and orange line) and $\protect\alpha N_F^{1/2}=0.36$
(red circles and dashed green line). } \label{FCS2}
\end{figure}
The relaxation to zero polarization in the strong coupling limit is a
non-trivial dynamical effect. In Eq.~(\ref{Pfinal}), we can separate
explicitly the time independent contributions coming from the states with $%
\epsilon_{\alpha}=\epsilon_{\beta}$ and write the polarization as $%
\mathcal{P}(t) = P_{\mathrm{const}} +\Delta P(t)$. For large but
finite values of the parameter $\alpha\sqrt{N_F}$ the
time-independent contribution $P_{\mathrm{const}}$ does not
vanish. Instead, after the initial relaxation, the time-dependent
term approaches a constant value, $\Delta P(t)\rightarrow
-P_{\mathrm{const}}$, leading to a vanishing polarization. The
time independent contribution $P_{\mathrm{const}}$ is shown in
Fig.~\ref{FCS2} for different sets of parameters ($\alpha, N_F$).
Notice that in the large coupling limit $P_{\mathrm{const}}$
depends on a single scaling parameter, $\alpha\sqrt{N_F}$. In
fact, as it is suggested by Eq.~(\ref{Pexpand}), the polarization
itself scales with $\alpha\sqrt{N_F}$ in the strong coupling limit
(inset of Fig.~\ref{FCS2}).

\begin{figure}[tbp]
\begin{center}
\includegraphics[width=0.42\textwidth]{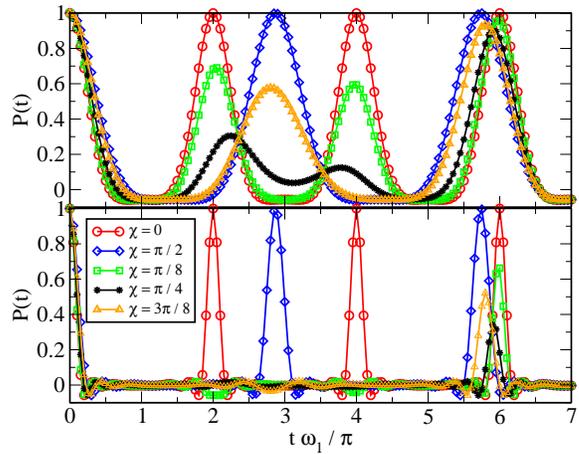}
\end{center}
\caption{(Color online) Spin polarization as a function of time in an
elliptic harmonic trap (with $\protect\omega_2/\protect\omega_1= 23/33$).
Relaxation curves are plotted for different laser orientations parameterized
by the angle $\protect\chi$ (see text). The upper and lower panels
correspond to the SO couplings $\protect\alpha =0.2$ and $\protect%
\alpha =0.6 $, respectively. For an arbitrary orientation of the lasers, $%
\mathcal{P}(t)$ has a period $23(2\protect\pi /\protect\omega _{2})=33(2%
\protect\pi /\protect\omega _{1})$. The residual echoes become
smaller and  vanish in the strong coupling limit. } \label{FCS3}
\end{figure}

The periodic time dependence of the polarization is a consequence
of the equidistant energy spectrum. What would be the result of
altering this structure?
We consider an elliptical trap described by the potential
$w(\mathbf{r})=m[\omega _{1}^{2}(x\cos \chi -y\sin \chi
)^{2}+\omega _{2}^{2}(x\sin \chi +y\cos \chi )^{2}]/2$, where the
angle $\chi$ describes the orientation of the symmetry axes of the
trap potential relative to the ($x, y$) directions defined by the
laser field. If $\chi =0$ ($\chi =\pi /2$), $\mathcal{P}(t)$ is
the same as for an isotropic system with a trap frequency $\omega
_{1}$ ($\omega _{2}$). If the orientation of the symmetry axes
deviates from the ($x, y$) directions, $\mathcal{P}(t)$ is still
periodic if $\omega _{2}/\omega _{1}$ is a rational number and
non-periodic otherwise. In the former case with $\omega
_{2}/\omega _{1} = p/q$, the period of oscillations is $T=p(2\pi
/\omega _{2})=q(2\pi /\omega _{1})$ (see, Fig.~\ref{FCS3}). Notice
that in the weak coupling regime (upper panel in Fig.~\ref{FCS3})
for small deviations from the symmetry axes, spin relaxation is
reminiscent of the periodic structure of the symmetric
configuration. However, the residual echoes become weaker
and eventually vanish in the strong coupling limit (lower panel in Fig.~\ref%
{FCS3}).


Finally, we briefly discuss how to observe spin relaxation and
echoes in ultracold Fermi gases through the time of flight
measurements. Initially the atoms are prepared in the state
$\left\vert 1\right\rangle $. To transfer them to the spin up
polarized state $\left\vert u_{1}\right\rangle $, one applies a
Raman pulse between the states $\left\vert 1\right\rangle $ and
$\left\vert 2\right\rangle $ with a spatially dependent Rabi
frequency that matches the spatial variation of the $\left\vert
u_{1}\right\rangle $ state. Then, the laser fields $L_{i0}$ are
turned on and the Fermi gas experiences the effective SO coupling.
After a certain time $t$, one turns off the laser fields $L_{i0}$
and applies a reversal Raman pulse to transfer the atoms from
$\left\vert u_{1}\right\rangle $ back to $\left\vert
1\right\rangle $. Finally, a time of flight measurement gives the
number of atoms in the $\left\vert 1\right\rangle $ state (i.e. in
$\left\vert
u_{1}\right\rangle $ at time t) and thus determines the spin polarization $%
\mathcal{P}(t)$. A possible choice of parameters is $\omega =2\pi \times
10Hz $, $\Delta =2\pi \times 10^{2}GHz$ , and $\Omega =2\pi \times 10^{7}Hz$%
, satisfying $\omega \ll \Omega \ll \Delta $ and $\omega \ll
\Omega ^{2}/\Delta $. The strength of the SO coupling
$v_{0}=v_{\phi }\cos \theta =\frac{2\hbar k_{1}}{m}\sin \left(
\frac{\zeta }{2}\right) \cos \theta $ can be varied from $0$ to
$\frac{2\hbar k_{1}}{m}\cos \theta $ by adjusting the angle $\zeta
$ between the laser beams. Consequently, $\alpha
=(mv_{0}^{2}/2\hbar \omega )^{1/2}=2\left( \omega _{r}/\omega
\right) ^{1/2}\sin \left( \frac{\zeta }{2}\right) \cos \theta $
(where $\omega _{r}$ is the photon recoil frequency for $L_{10}$),
can vary in a range from $0$ to $100$ for $^{6}$Li atoms.
Therefore, both the strong and weak SO coupling limits are
accessible in experiment.


C.Z. was supported by ARO-DTO, ARO-LPS, and LPS-NSA.

\vspace*{-0.1in}
\bibliography{coldspin}

\begin{thebibliography}{20}
\expandafter\ifx\csname natexlab\endcsname\relax\def\natexlab#1{#1}\fi
\expandafter\ifx\csname bibnamefont\endcsname\relax
  \def\bibnamefont#1{#1}\fi
\expandafter\ifx\csname bibfnamefont\endcsname\relax
  \def\bibfnamefont#1{#1}\fi
\expandafter\ifx\csname citenamefont\endcsname\relax
  \def\citenamefont#1{#1}\fi
\expandafter\ifx\csname url\endcsname\relax
  \def\url#1{\texttt{#1}}\fi
\expandafter\ifx\csname urlprefix\endcsname\relax\def\urlprefix{URL }\fi
\providecommand{\bibinfo}[2]{#2}
\providecommand{\eprint}[2][]{\url{#2}}

\bibitem[{\citenamefont{Lewenstein et~al.}()\citenamefont{Lewenstein, Sanpera,
  Ahufinger, Damski, {Sen~De}, and Sen}}]{Lewenstein}
\bibinfo{author}{\bibfnamefont{M.}~\bibnamefont{Lewenstein}},
  \bibinfo{author}{\bibfnamefont{A.}~\bibnamefont{Sanpera}},
  \bibinfo{author}{\bibfnamefont{V.}~\bibnamefont{Ahufinger}},
  \bibinfo{author}{\bibfnamefont{B.}~\bibnamefont{Damski}},
  \bibinfo{author}{\bibfnamefont{A.}~\bibnamefont{{Sen~De}}}, \bibnamefont{and}
  \bibinfo{author}{\bibfnamefont{U.}~\bibnamefont{Sen}},
  \bibinfo{howpublished}{cond-mat/0606771.}

\bibitem[{\citenamefont{Juzeliunas and {O}hberg}(2004)}]{oberg1}
\bibinfo{author}{\bibfnamefont{G.}~\bibnamefont{Juzeliunas}} \bibnamefont{and}
  \bibinfo{author}{\bibfnamefont{P.}~\bibnamefont{{O}hberg}},
  \bibinfo{journal}{\prl} \textbf{\bibinfo{volume}{93}},
  \bibinfo{pages}{033602} (\bibinfo{year}{2004}).

\bibitem[{\citenamefont{Juzeliunas et~al.}(2005)\citenamefont{Juzeliunas,
  {O}hberg, Ruseckas, and Klein}}]{oberg2}
\bibinfo{author}{\bibfnamefont{G.}~\bibnamefont{Juzeliunas}},
  \bibinfo{author}{\bibfnamefont{P.}~\bibnamefont{{O}hberg}},
  \bibinfo{author}{\bibfnamefont{J.}~\bibnamefont{Ruseckas}}, \bibnamefont{and}
  \bibinfo{author}{\bibfnamefont{A.}~\bibnamefont{Klein}},
  \bibinfo{journal}{\pra} \textbf{\bibinfo{volume}{71}},
  \bibinfo{pages}{053614} (\bibinfo{year}{2005}).

\bibitem[{\citenamefont{Jaksch and Zoller}(2003)}]{Jaksch}
\bibinfo{author}{\bibfnamefont{D.}~\bibnamefont{Jaksch}} \bibnamefont{and}
  \bibinfo{author}{\bibfnamefont{P.}~\bibnamefont{Zoller}},
  \bibinfo{journal}{New J. Phys.} \textbf{\bibinfo{volume}{5}},
  \bibinfo{pages}{56} (\bibinfo{year}{2003}).

\bibitem[{\citenamefont{Ruseckas et~al.}(2005)\citenamefont{Ruseckas,
  Juzeliunas, {O}hberg, and Fleischhauer}}]{Rusec}
\bibinfo{author}{\bibfnamefont{J.}~\bibnamefont{Ruseckas}},
  \bibinfo{author}{\bibfnamefont{G.}~\bibnamefont{Juzeliunas}},
  \bibinfo{author}{\bibfnamefont{P.}~\bibnamefont{{O}hberg}}, \bibnamefont{and}
  \bibinfo{author}{\bibfnamefont{M.}~\bibnamefont{Fleischhauer}},
  \bibinfo{journal}{\prl} \textbf{\bibinfo{volume}{95}},
  \bibinfo{pages}{010404} (\bibinfo{year}{2005}).

\bibitem[{\citenamefont{Osterloh et~al.}(2005)\citenamefont{Osterloh, Baig,
  Santos, Zoller, and Lewenstein}}]{Osterloh}
\bibinfo{author}{\bibfnamefont{K.}~\bibnamefont{Osterloh}},
  \bibinfo{author}{\bibfnamefont{M.}~\bibnamefont{Baig}},
  \bibinfo{author}{\bibfnamefont{L.}~\bibnamefont{Santos}},
  \bibinfo{author}{\bibfnamefont{P.}~\bibnamefont{Zoller}}, \bibnamefont{and}
  \bibinfo{author}{\bibfnamefont{M.}~\bibnamefont{Lewenstein}},
  \bibinfo{journal}{\prl} \textbf{\bibinfo{volume}{95}},
  \bibinfo{pages}{010403} (\bibinfo{year}{2005}).

\bibitem[{\citenamefont{Zhu et~al.}(2006)\citenamefont{Zhu, Fu, Wu, Zhang, and
  Duan}}]{Zhu}
\bibinfo{author}{\bibfnamefont{S.}~\bibnamefont{Zhu}},
  \bibinfo{author}{\bibfnamefont{H.}~\bibnamefont{Fu}},
  \bibinfo{author}{\bibfnamefont{C.}~\bibnamefont{Wu}},
  \bibinfo{author}{\bibfnamefont{S.}~\bibnamefont{Zhang}}, \bibnamefont{and}
  \bibinfo{author}{\bibfnamefont{L.}~\bibnamefont{Duan}},
  \bibinfo{journal}{\prl} \textbf{\bibinfo{volume}{97}},
  \bibinfo{pages}{240401} (\bibinfo{year}{2006}).

\bibitem[{\citenamefont{Satija et~al.}(2006)\citenamefont{Satija, Dakin, and
  Clark}}]{Clark}
\bibinfo{author}{\bibfnamefont{I.~I.} \bibnamefont{Satija}},
  \bibinfo{author}{\bibfnamefont{D.~C.} \bibnamefont{Dakin}}, \bibnamefont{and}
  \bibinfo{author}{\bibfnamefont{C.~W.} \bibnamefont{Clark}},
  \bibinfo{journal}{\prl} \textbf{\bibinfo{volume}{97}},
  \bibinfo{pages}{216401} (\bibinfo{year}{2006}).

\bibitem[{\citenamefont{Dyakonov and Perel}(1971)}]{DyakPer}
\bibinfo{author}{\bibfnamefont{M.~I.} \bibnamefont{Dyakonov}} \bibnamefont{and}
  \bibinfo{author}{\bibfnamefont{V.~I.} \bibnamefont{Perel}},
  \bibinfo{journal}{Sov. Phys. - JETP} \textbf{\bibinfo{volume}{33}},
  \bibinfo{pages}{1053} (\bibinfo{year}{1971}).

\bibitem[{\citenamefont{Khaetskii and Nazarov}(2000)}]{Khaets}
\bibinfo{author}{\bibfnamefont{A.~V.} \bibnamefont{Khaetskii}}
  \bibnamefont{and} \bibinfo{author}{\bibfnamefont{Y.~V.}
  \bibnamefont{Nazarov}}, \bibinfo{journal}{\prb}
  \textbf{\bibinfo{volume}{61}}, \bibinfo{pages}{12639} (\bibinfo{year}{2000}).

\bibitem[{\citenamefont{Woods et~al.}(2002)\citenamefont{Woods, Reinecke, and
  {Lyanda-Geller}}}]{Woods}
\bibinfo{author}{\bibfnamefont{L.}~\bibnamefont{Woods}},
  \bibinfo{author}{\bibfnamefont{T.}~\bibnamefont{Reinecke}}, \bibnamefont{and}
  \bibinfo{author}{\bibfnamefont{Y.}~\bibnamefont{{Lyanda-Geller}}},
  \bibinfo{journal}{\prb} \textbf{\bibinfo{volume}{66}},
  \bibinfo{pages}{161318} (\bibinfo{year}{2002}).

\bibitem[{\citenamefont{Merkulov et~al.}(2002)\citenamefont{Merkulov, Efros,
  and Rosen}}]{Merkul}
\bibinfo{author}{\bibfnamefont{I.}~\bibnamefont{Merkulov}},
  \bibinfo{author}{\bibfnamefont{A.}~\bibnamefont{Efros}}, \bibnamefont{and}
  \bibinfo{author}{\bibfnamefont{M.}~\bibnamefont{Rosen}},
  \bibinfo{journal}{\prb} \textbf{\bibinfo{volume}{65}},
  \bibinfo{pages}{205309} (\bibinfo{year}{2002}).

\bibitem[{\citenamefont{Semenov and Kim}(2004)}]{Semen}
\bibinfo{author}{\bibfnamefont{Y.}~\bibnamefont{Semenov}} \bibnamefont{and}
  \bibinfo{author}{\bibfnamefont{K.}~\bibnamefont{Kim}},
  \bibinfo{journal}{\prl} \textbf{\bibinfo{volume}{92}},
  \bibinfo{pages}{026601} (\bibinfo{year}{2004}).

\bibitem[{\citenamefont{Chang et~al.}(2004)\citenamefont{Chang, Mal'shukov, and
  Chao}}]{Chang}
\bibinfo{author}{\bibfnamefont{C.}~\bibnamefont{Chang}},
  \bibinfo{author}{\bibfnamefont{A.}~\bibnamefont{Mal'shukov}},
  \bibnamefont{and} \bibinfo{author}{\bibfnamefont{K.}~\bibnamefont{Chao}},
  \bibinfo{journal}{\prb} \textbf{\bibinfo{volume}{70}},
  \bibinfo{pages}{245309} (\bibinfo{year}{2004}).

\bibitem[{\citenamefont{Milner et~al.}(2001)\citenamefont{Milner, Hanssen,
  Campbell, and Raizen}}]{Milner}
\bibinfo{author}{\bibfnamefont{V.}~\bibnamefont{Milner}},
  \bibinfo{author}{\bibfnamefont{J.}~\bibnamefont{Hanssen}},
  \bibinfo{author}{\bibfnamefont{W.}~\bibnamefont{Campbell}}, \bibnamefont{and}
  \bibinfo{author}{\bibfnamefont{M.}~\bibnamefont{Raizen}},
  \bibinfo{journal}{Phys. Rev. Lett.} \textbf{\bibinfo{volume}{86}},
  \bibinfo{pages}{1514} (\bibinfo{year}{2001}).

\bibitem[{\citenamefont{Friedman et~al.}(2001)\citenamefont{Friedman, Kaplan,
  Carasso, and Davidson}}]{Friedman}
\bibinfo{author}{\bibfnamefont{N.}~\bibnamefont{Friedman}},
  \bibinfo{author}{\bibfnamefont{A.}~\bibnamefont{Kaplan}},
  \bibinfo{author}{\bibfnamefont{D.}~\bibnamefont{Carasso}}, \bibnamefont{and}
  \bibinfo{author}{\bibfnamefont{N.}~\bibnamefont{Davidson}},
  \bibinfo{journal}{Phys. Rev. Lett.} \textbf{\bibinfo{volume}{86}},
  \bibinfo{pages}{1518} (\bibinfo{year}{2001}).

\bibitem[{\citenamefont{Meyrath et~al.}(2005)\citenamefont{Meyrath, Schreck,
  Hanssen, Chuu, and Raizen}}]{Raizen}
\bibinfo{author}{\bibfnamefont{T.~P.} \bibnamefont{Meyrath}},
  \bibinfo{author}{\bibfnamefont{F.}~\bibnamefont{Schreck}},
  \bibinfo{author}{\bibfnamefont{J.~L.} \bibnamefont{Hanssen}},
  \bibinfo{author}{\bibfnamefont{C.}~\bibnamefont{Chuu}}, \bibnamefont{and}
  \bibinfo{author}{\bibfnamefont{M.}~\bibnamefont{Raizen}},
  \bibinfo{journal}{\pra} \textbf{\bibinfo{volume}{71}},
  \bibinfo{pages}{041604} (\bibinfo{year}{2005}).

\bibitem[{\citenamefont{Spielman et~al.}(2007)\citenamefont{Spielman, Phillips,
  and Porto}}]{Spielman}
\bibinfo{author}{\bibfnamefont{I.}~\bibnamefont{Spielman}},
  \bibinfo{author}{\bibfnamefont{W.}~\bibnamefont{Phillips}}, \bibnamefont{and}
  \bibinfo{author}{\bibfnamefont{J.}~\bibnamefont{Porto}},
  \bibinfo{journal}{Phys. Rev. Lett.} \textbf{\bibinfo{volume}{98}},
  \bibinfo{pages}{080404} (\bibinfo{year}{2007}).

\bibitem[{\citenamefont{Grimm et~al.}(2000)\citenamefont{Grimm, Weidemuller,
  and Ovchinnikov}}]{Grimm}
\bibinfo{author}{\bibfnamefont{R.}~\bibnamefont{Grimm}},
  \bibinfo{author}{\bibfnamefont{M.}~\bibnamefont{Weidemuller}},
  \bibnamefont{and}
  \bibinfo{author}{\bibfnamefont{Y.}~\bibnamefont{Ovchinnikov}},
  \bibinfo{journal}{Adv. At. Mol. Opt. Phys} \textbf{\bibinfo{volume}{42}},
  \bibinfo{pages}{95} (\bibinfo{year}{2000}).

\bibitem[{\citenamefont{Mishchenko et~al.}()\citenamefont{Mishchenko, Shytov,
  and Halperin}}]{CMSO}
\bibinfo{author}{\bibfnamefont{E.~G.} \bibnamefont{Mishchenko}},
  \bibinfo{author}{\bibfnamefont{A.~V.} \bibnamefont{Shytov}},
  \bibnamefont{and} \bibinfo{author}{\bibfnamefont{B.~I.}
  \bibnamefont{Halperin}}, \bibinfo{howpublished}{Phys. Rev. Lett. {\bf 93},
  226602 (2004); V.~M.~Galitski, A.~A.~Burkov, and S.~{Das~Sarma}, Phys. Rev. B
  {\bf 74}, 115331 (2006).}

\end{thebibliography}

\end{document}